\documentclass[12pt,amsfonts]{article}
\usepackage{amsmath,amsfonts,latexsym}
\usepackage{color}
\usepackage[]{hyperref}

\parskip=\medskipamount

\renewcommand{\a}{\alpha}
\renewcommand{\b}{\beta}
\renewcommand{\c}{\chi}
\renewcommand{\d}{\delta}
\newcommand{\e}{\epsilon}
\newcommand{\ce}{\varepsilon}

\newcommand{\g}{\gamma}
\newcommand{\h}{\eta}

\newcommand{\m}{\mu}

\newcommand{\p}{\pi}
\newcommand{\q}{\theta}

\newcommand{\s}{\sigma}

\newcommand{\w}{\omega}

\newcommand{\z}{\zeta}
\newcommand{\D}{\Delta}
\newcommand{\F}{\Phi}
\newcommand{\G}{\Gamma}

\newcommand{\Y}{\Psi}

\newcommand{\ex}{\mathrm{e}}
\newcommand{\ba}{\bar}
\renewcommand{\.}{\dot}
\newcommand{\pd}{\partial}
\renewcommand{\~}{\tilde}
\newcommand{\nab}{\nabla}
\newcommand{\ii}{\mathrm{i}}
\newcommand{\dd}{\mathrm{d}}
\newcommand{\inte}{\int\!\!\dd}
\newcommand{\no}{\nonumber}
\newcommand{\gc}[2]{\left. \lbrack #1, #2\} \right.}

\newcommand{\cd}{\mathcal{D}}
\newcommand{\tr}{\textrm{tr}}
\newcommand{\Ad}{\textrm{Ad}}

\newcommand{\1}[1][\a]{W^{#1}}
\newcommand{\2}[1][\a]{W_{#1}}
\newcommand{\3}[1][\a]{\ba{W}_{\.{#1}}}
\newcommand{\4}[1][\a]{\ba{W}^{\.{#1}}}
\newcommand{\5}{W^2}
\newcommand{\6}{\ba{W}^2}

\newcommand{\FAA}{F^{ab} F_{ab} F^{cd} F_{cd}}
\newcommand{\FBB}{F^{ab} F^{cd} F_{ab} F_{cd}}
\newcommand{\FCC}{F^{ab} F_{bc} F^{cd} F_{da}}
\newcommand{\FEE}{F^{ab} F_{bc} F_{ad} F^{dc}}

\newcommand{\FA}{F^{ab}F_{bc}F^{cd}F_{de}F^{e}_{\phantom{e} a}}
\newcommand{\FB}{F^{ab}F_{bc}F^{cd}F^{e}_{\phantom{e} a} F_{de}}
\newcommand{\FC}{F^{ab}F^{cd}F_{bc}F^{e}_{\phantom{e} a} F_{de}}
\newcommand{\FD}{F^{ab}F^{cd}F^{e}_{\phantom{e} a} F_{bc}F_{de}}
\newcommand{\FE}{F^{ab}F_{bc}F^{c}_{\phantom{e} a} F^{de}F_{de}}
\newcommand{\FF}{F^{ab}F^{de}F_{bc}F^{c}_{\phantom{e} a} F_{de}}

\newcommand{\Fa}{(\nab^{e}F^{ab})(\nab_{e}F_{ab})F^{cd}F_{cd}}
\newcommand{\Fb}{(\nab^e F^{ab})(\nab_e F^{cd}) F_{ab} F_{cd}}
\newcommand{\Fc}{(\nab^e F^{ab})(\nab_e F^{cd}) F_{cd} F_{ab}}
\newcommand{\Fd}{(\nab^e F^{ab})(\nab_e F_{bc}) F^{cd} F_{da}}
\newcommand{\Fe}{(\nab^e F^{ab})(\nab_e F_{ca}) F_{bd} F^{dc}}
\newcommand{\Ff}{(\nab^e F^{ab})(\nab_e F^{cd}) F_{bc} F_{da}}
\newcommand{\Fg}{(\nab^e F^{ab}) F_{da}(\nab_e F_{bc}) F^{cd}}
\newcommand{\Fh}{(\nab^e F^{ab}) F^{cd}(\nab_e F_{ab}) F_{cd}}
\newcommand{\Fi}{(\nab^e F^{ab}) F_{bc}(\nab_e F^{cd}) F_{da}}
\newcommand{\Fj}{(\nab^e F^{ab}) F_{ab}(\nab_e F^{cd}) F_{cd}}
\newcommand{\Fk}{(\nab^e F^{ab}) F^{cd}(\nab_e F_{cd}) F_{ab}}
\newcommand{\Fl}{(\nab^e F^{ab}) F^{dc}(\nab_e F_{cb}) F_{da}}

\newcommand{\fb}{F^{ab}(\nab_a F^{cd})(\nab^e F_{bc}) F_{de}}

\newcommand{\fd}{(\nab^a F^{ef})(\nab_b F_{ef})F^{ac} F_{cb}}
\newcommand{\fe}{(\nab^a F^{ef})(\nab_b F_{ef})F^{bc} F_{ca}}
\newcommand{\ff}{(\nab^a F^{ef})F^{cb}(\nab_b F_{ef}) F_{ac}}
\newcommand{\fg}{(\nab_b F^{ef})F^{cb}(\nab^a F_{ef}) F_{ac}}
\newcommand{\fh}{(\nab^b F^{cd})(\nab_c F^{ea})F_{de} F_{ab}}
\newcommand{\fj}{(\nab_e F^{bc})(\nab^a F^{de})F_{ab} F_{cd}}
\newcommand{\fk}{(\nab^a F^{de})(\nab^b F_{ec})F_{ab} F^{c}_{\phantom{c} d}}
\newcommand{\fl}{(\nab_a F^{de})(\nab^c F_{be})F^{ab} F_{cd}}
\newcommand{\fm}{(\nab^a F^{de})(\nab_e F^{cb})F_{ab} F_{cd}}
\newcommand{\fn}{(\nab_e F^{bc})F_{cd}(\nab^d F^{ea}) F_{ab}}
\newcommand{\fo}{(\nab^a F^{cd})(\nab^b F_{de})F^{e}_{\phantom{e}c} F_{ab}}
\newcommand{\fp}{(\nab_a F^{cd})(\nab^e F^{ab})F_{cb} F_{ed}}
\newcommand{\fq}{F^{de}(\nab_e F^{cb})F_{ab} (\nab^a F_{cd})}
\newcommand{\fr}{F_{ab}(\nab_e F^{bc})F^{ed} (\nab^a F_{dc})}
\newcommand{\fs}{F_{ab}(\nab_e F^{bc})F_{cd} (\nab^a F^{de})}

\begin{document}

\begin{titlepage}
\thispagestyle{empty}

\begin{flushright}
\end{flushright}

\vspace{5mm}
\begin{center}
\Large{\textbf{Higher order contributions to the effective action
of $\mathcal{N}=2$ super Yang-Mills}}
\end{center}
\vspace{3mm}

\begin{center}
\large{Darren T. Grasso}\\
\vspace{2mm}

\footnotesize{\textit{School of Physics, The University of Western
Australia\\ Crawley, W.A. 6009. Australia}} \\ \vspace{3mm}
\href{mailto:grasso@physics.uwa.edu.au}{\tt{grasso@physics.uwa.edu.au}}

\vspace{2mm}

\end{center}
\vspace{5mm}

\begin{abstract}
\noindent We apply heat kernel techniques in $\mathcal{N}=1$
superspace to compute the one-loop effective action to order $F^5$
for chiral superfields coupled to a non-Abelian super Yang-Mills
background. The results, when combined with those of
hep-th/0210146, yield the one-loop effective action to order $F^5$
for any $\mathcal{N}=2$ super Yang-Mills theory coupled to matter
hypermultiplets.
\end{abstract}

\vfill
\end{titlepage}

\newpage

\setcounter{page}{1} \setcounter{equation}{0}
\section{Introduction}
\label{intro}

The low energy effective actions for D-branes in superstring
theories are supersymmetric Yang-Mills theories. This fact has led
to a remarkably fruitful interface between superstring theory and
supersymmetric Yang-Mills theory, which has yielded valuable
insights on both sides. Perhaps the most spectacular example is
the Maldacena conjecture of the duality between $\mathcal{N}=4$
supersymmetric Yang-Mills theory and type IIB superstring theory
in an $AdS_5 \times S^5$ background
\cite{Mald,Gubser:1998bc,Witten:1998qj}. On the field theory side,
more stringent tests of this conjecture require the computation of
the effective action of $\mathcal{N}=4$ supersymmetric Yang-Mills
theory. In \cite{grasso}, the one-loop effective action for
$\mathcal{N}=4$ supersymmetric Yang-Mills theory was computed
through to order $F^6$ using the $\mathcal{N}=1$ superfield
formulation. At the component level the results to order $F^5$
were found to be in agreement with string theory calculations to
the same order \cite{Koerber} (see also \cite{Refolli:2001df}).

Maldacena's original arguments have been generalized to include a
conjectured duality between certain $\mathcal{N}=2$ superconformal
supersymmetric Yang-Mills theories and superstring theories in
special backgrounds \cite{Aharony,Douglas}. As in the
$\mathcal{N}=4$ case, more detailed tests of these conjectures
will require comparison of the effective action for specific
$\mathcal{N}=2$ supersymmetric Yang-Mills theories with results
derived from superstring calculations. In anticipation of such
tests, it is important that the effective action of these
supersymmetric Yang-Mills theories be computed to higher orders.
This paper details the computation of the one-loop effective
action for general $\mathcal{N}=2$ supersymmetric Yang-Mills
theories to order $F^5$ in $\mathcal{N}=1$ superfield formulation.
Since at the quantum level, with the same gauge conditions,
$\mathcal{N}=2$ and $\mathcal{N}=4$ super Yang-Mills theories
differ only in the hypermultiplet sector, making use of the
results already derived in \cite{grasso}, we need only calculate
the one-loop contribution associated with this sector.

This paper is organized as follows. Firstly, in Section
\ref{sec:yangmills}, we briefly discuss the connection between
$\mathcal{N}=2$ and $\mathcal{N}=4$ super Yang-Mills theories at
the classical and one-loop quantum level. In Section
\ref{sec:oneloop} we review, using heat kernels and zeta
functional regularization, the quantization of matter coupled to a
non-Abelian super Yang-Mills background in $\mathcal{N}=1$
superspace. We also demonstrate how the one-loop effective action
may be cast on full superspace, rather than a chiral subspace.
Section \ref{sec:methodofcomputation} briefly describes the method
of computation of the associated heat kernel, being further
explained in Appendix A. The results of the calculation in full
superfield form, as well as its leading bosonic component, are
given to order $F^5$ in Section \ref{sec:results}. Section
\ref{sec:comparison} compares these results with partial bosonic
results currently available in the literature, and Section
\ref{sec:discussion} discusses the one-loop effective action for
general $\mathcal{N}=2$ super Yang-Mills theory to this order.
Appendix B details the change of basis calculations required to
compare the $F^5$ component results with literature.

We adopt the
conventions and notation of \cite{Wess} and \cite{kuzenko}.

\setcounter{equation}{0}
\section{$\mathcal{N}=2$ super
Yang-Mills} \label{sec:yangmills}

The most general classical non-Abelian $\mathcal{N}=2$
supersymmetric Yang-Mills action cast in $\mathcal{N}=1$
superfield form consists of two parts,
\begin{equation}\label{eq:n2action}
S_{\mathcal{N}=2}=S_\textrm{pure}+S_{\textrm{hyper}}.
\end{equation}
The pure $\mathcal{N}=2$ super Yang-Mills action is given by:
\begin{equation}
S_\textrm{pure}=\frac{1}{g^2}\tr_\Ad\left(\inte^8z \; \F^{\dag}\F
+\frac{1}{2}\inte^6 z \; \5 \right),
\end{equation}
where Ad denotes the adjoint representation.  The hypermultiplet
action is
\begin{multline}\label{eq:hyperaction}
S_{\textrm{hyper}}=\inte^8z
\big(Q^{\dag}Q+\~{Q}^\dag\~{Q}\big)+\sqrt{2}\inte^6 z \;
\~{Q}^{\textrm{T}}\F \,Q + \sqrt{2}\inte^6 \ba{z} \;
Q^{\dag}\F^{\dag}\ba{\~Q}\\+ \mathcal{M}\inte^6 z \;
\~{Q}^{\textrm{T}} \,Q + \mathcal{M} \inte^6 \ba{z} \;
Q^{\dag}\ba{\~Q}
\end{multline}
where $\mathcal{M}$ is the hypermultiplet mass. The covariantly
chiral superfields $Q$, $\~Q$ and $\F$ transform in the
representations $R$, its conjugate $R_c$, and the adjoint
representation of the gauge group respectively. $\1$ is the
covariantly chiral superfield strength associated with the gauge
covariant derivatives,
\begin{gather}
\cd_A=(\cd_a,\cd_\a,\ba{\cd}^{\.\a})=D_A+\ii\G_A, \qquad
\G_A=\G^I_A(z) T^I, \\
(T^I)^\dag=T^I, \qquad\qquad [T^I,T^J]=\ii f^{IJK}T^K,
\end{gather}
where $D_A$ are flat covariant derivatives and $T^I$ are the
Hermitian generators of the gauge group.  The gauge covariant
derivatives satisfy the algebra:
\begin{gather}
\{\cd_\a,\cd_\b\}=\{\ba{\cd}_{\.{\a}},\ba{\cd}_{\.{\b}}\}=0 \no \\
\{\cd_\a,\ba{\cd}_{\.{\a}}\}=-2\ii \cd_{\a \.{\a}}=-2\ii
(\s^a)_{\a\.{\a}}\cd_a \no \\
[\cd_\a, \cd_{\b\.{\b}}]=2\ii \ce_{\a \b}\ba{W}_{\.{\b}}
\label{eq:commutation} \\
[\ba{\cd}_{\.{\a}},\cd_{\b\.{\b}}]=2\ii \ce_{\.{\a}\.{ \b}}W_{\b}
\no \\
[\cd_{\a\.{\a}},\cd_{\b\.{\b}}]=(\s^a)_{\a\.{\a}}(\s^b)_{\b\.{\b}}G_{ab}
=-\ce_{\a\b}(\ba{\cd}_{\.\a}\ba{W}_{\.\b})-\ce_{\.{\a}\.{\b}}(\cd_\a{W}_\b).\no
\end{gather}

We are interested in the one-loop effective action for a generic
$\mathcal{N}=2$ SYM theory (\ref{eq:n2action}), which can be
computed efficiently using the $\mathcal{N}=1$ background field
formalism \cite{GGRS,Grisaru:1979wc}.  Here we consider providing
only the $\mathcal{N}=1$ vector multiplet with background field
values. After background covariant gauge fixing
\cite{Grisaru:1979wc,Grisaru:1981xm}, the effective action at the
one-loop level is a linear combination of two types of
contribution \cite{GGRS,Banin:2002mf} - the logarithm of the
superdeterminant of the background covariant operator in the
quadratic part of the action for a vector superfield in the
adjoint representation, and the logarithm of the superdeterminant
of the operator which results from the quadratic part of the
action for a  background covariantly chiral scalar superfield in a
real representation of the gauge group (either the adjoint
representation or $R \oplus R_c$). Since the former has already
been dealt with in \cite{grasso}, it remains to compute the latter
in order to be able to assemble the one-loop low energy effective
action for an arbitrary $\mathcal{N}=2$ SYM theory. For related
discussions, material and different approaches to analogous
problems also see
\cite{add1,add2,add3,add4,add5,add6,add7,add8,add9,add10,Banin:2003es}.

\setcounter{equation}{0}
\section{Chiral superfields in a super Yang-Mills
background} \label{sec:oneloop}

As discussed above, the problem of computing the one-loop
effective action of any $\mathcal{N}=2$ SYM theory is reduced to
computing the effective action for a background covariantly chiral
scalar $\c=\{\c^i(z)\}$, which transforms under some real
representation $\mathcal{R}$ of the gauge group. The background
covariant derivatives are given by
\begin{gather}
\cd_A=(\cd_a,\cd_\a,\ba{\cd}^{\.\a})=D_A+\ii\G_A, \qquad
\G_A=\G^I_A(z) \~T^I,
\end{gather}
and satisfy the algebra (\ref{eq:commutation}), where here the
Hermitian generators $(\~T^I)^i_{\phantom{i}j}$ are antisymmetric:
\begin{equation}
(\~T^I)^\dag=\~T^I \qquad\qquad (\~T^I)^{\textrm{T}}=-\~T^I,
\end{equation}\enlargethispage{\baselineskip}
Since we are interested only in one-loop corrections,  any
interactions may be ignored, and the classical action in question
is:
\begin{equation}\label{eq:action}
S[\c,\ba{\c}]=\inte^8z\; \c^\dag\c +\frac{m}{2}\bigg\{\inte^6 z\;
\c^{\mathrm{T}}\c +c.c. \bigg\}.
\end{equation}
The mass $m$ is either an explicit mass (as in the case of the
hypermultiplet mass $\mathcal{M}$ in (\ref{eq:hyperaction})), or a
infrared regulator for massless chiral scalars (as in the case of
$\F$ in (\ref{eq:hyperaction}) and the ghosts)\footnote{One may
also introduce a background for the adjoint scalar $\F$ in
(\ref{eq:n2action}) taking values in the Cartan subalgebra, which
then generates masses for the vector multiplet, the hypermultiplet
and ghosts.  For a discussion see \cite{Banin:2003es}.}.

One can efficiently calculate the one-loop contribution to the
effective action via zeta function regularization, and it is
proportional to $\z'(0)$, where the zeta function is defined by
\begin{equation}
\z(s)=\frac{\m^{2s}}{\G(s)}\int_0^\infty\!\!\dd t \; t^{s-1} e^{-t
m^2}\big( K_{+}(t)+K_{-}(t)\big).
\end{equation}
In this expression, $\m$ is the renormalization point and $K_+(t)$
and $K_-(t)$ are the functional traces of the chiral and
antichiral heat kernels respectively, which are defined by:
\begin{eqnarray}
K_{\pm}(t)= \tr_{\mathcal{R}}\inte^6 z_{\pm} \inte^6 z_{\pm}'
\d_{\pm}(z,z') \ e^{t \Box_{\pm}} \d_{\pm}(z,z') \equiv
\tr_{\mathcal{R}}\inte^6 z_\pm \,K_\pm(z,t).
\end{eqnarray}
Here $\tr_{\mathcal{R}}$ denotes the trace over the representation
$\mathcal{R}$, $\dd z_\pm$ the integration measure over
(anti)chiral subspace, $\d_{\pm}(z,z')$ the (anti)chiral delta
functions,
\begin{gather}
\d_{+}(z,z')=-\frac{1}{4}\, \ba{\cd}^2 \mathbb{I} \d^{(8)}(z,z') \\
\d_{-}(z,z')=-\frac{1}{4} \,  \cd^2 \mathbb{I} \d^{(8)}(z,z') \\
\d^{(8)}(z,z')=\d^{(4)}(x,x')\d^{(2)}(\q-\q')\d^{(2)}(\ba{\q}-\ba{\q}'),
\end{gather}
and
\begin{align}\label{eq:operator}
\Box_{+}&=\frac{1}{16}\ba{\cd}^2 \cd^2\\
\Box_{-}&=\frac{1}{16}\cd^2\ba{\cd}^2.
\end{align}
It can be shown that
\begin{equation}
K_+(t)=K_{-}(t)
\end{equation}
and so
\begin{equation}\label{eq:zeta}
\z(s)=\frac{2 \m^{2s}}{\G(s)}\int_0^\infty\!\!\dd t \; t^{s-1}
e^{-t m^2}K_{+}(t),
\end{equation}
requiring computation of only the chiral kernel.

The operator
\begin{equation}
\Box_+=\frac{1}{16}\ba{\cd}^2\cd^2,
\end{equation}
is a Laplace-type operator when acting on chiral superfields,
$\Y$,
\begin{equation}
\Box_+ \, \Y = (\cd^a \cd_a -W^{\a}\cd_{\a} - \frac12 (\cd^{\a}
W_{\a}) ) \Y.
\end{equation}
This can be established using the identity
\begin{equation}\label{eq:iddd}
\cd^a \cd_a -W^{\a}\cd_{\a} =-\frac{1}{8}\ba{\cd}_{\.{\a}}
\cd^2\ba{\cd}^{\.{\a}}+\frac{1}{2}(\cd^\a W_\a)
+\frac{1}{16}(\cd^2 \ba{\cd}^2+\ba{\cd}^2 \cd^2).
\end{equation}
To compute the effective action, it suffices to consider an on-shell
background, $\cd^\a W_\a=\ba{\cd}_{\.{\a}}
\ba{W}^{\.{\a}}=0$, so that
\begin{equation}
\Box_+=\frac{1}{16}\ba{\cd}^2 \cd^2=\cd^a \cd_a -W^{\a}\cd_{\a}
\end{equation}
acting on chiral superfields.

$K_+(z,t)$ has an asymptotic expansion in $t$ in the limit
$t\rightarrow 0$, which is usually expressed
\begin{equation}\label{eq:asymptotica}
K_+(z,t)=\frac{1}{16 \p^2 t^2}\sum_{n=0}^{\infty}t^{n} a_n (z),
\qquad a_0 = a_1 = 0.
\end{equation}
The $a_n(z)$ are the DeWitt-Seeley coefficients, and are chiral
superfields which at the component level contain bosonic field
strength terms of the form $F^n$. To the best of our knowledge
only the first non-trivial coefficient $a_2$ is known in the
non-Abelian case \cite{buchbinder}:
\begin{equation}
a_2= W^2.
\end{equation}

Evaluating the one-loop effective action,
$\G^{(1)}_{\c,\mathcal{R}}$,  therefore amounts to computing the
DeWitt-Seeley coefficients:
\begin{equation}\label{eq:effectiveaction}
\G^{(1)}_{\c,\mathcal{R}} = \frac{1}{4}\z'(0) = \frac{1}{16 \p^2}
\ln{\left(\frac{\m}{m}\right)}\inte^{6}z \, \tr_{\mathcal{R}}(a_2)
+ \frac{1}{32 \p^2} \sum_{n=3}^\infty
\frac{(n-3)!}{m^{2n-4}}\inte^{6}z \, \tr_{\mathcal{R}}(a_n).
\end{equation}

It turns out that the $a_n$ with $n\geq3$ are expressible as
$\ba{\cd}^2$ acting on field strengths and their covariant
derivatives, and so the second term on the right hand side of
(\ref{eq:effectiveaction}) can be lifted to a gauge-invariant
superfunctional on full superspace. This can be proven as follows.

By differentiating the kernel $K_+(z,t)$ with respect to $t$, one
observes that:
\begin{align}
\frac{d K_+(z,t)}{d t}&=\frac{1}{16}\inte ^6 z_+'
\d_{+}(z,z')\ba{\cd}^2 \cd^2\ex^{t\Box_+}\d_{+}(z,z') \no \\
&=\frac{1}{16}\ba{\cd}^2\left(\inte ^6 z_+' \d_{+}(z,z')
\cd^2\ex^{t\Box_+}\d_{+}(z,z')\right) \no \\
&=\frac{1}{16}\ba{\cd}^2\left(\lim_{z'\rightarrow z}
\cd^2\ex^{t\Box_+}\d_{+}(z,z')\right), \label{eq:dk}
\end{align}
since
\begin{equation}
\ba{\cd}_{\.{\a}}\d_{+}(z,z')=0.
\end{equation}
On the other hand, (\ref{eq:asymptotica}) yields
\begin{equation}\label{eq:asymptoticda}
\frac{d K_+(z,t)}{d t}=\frac{1}{16
\p^2}\sum_{n=3}^{\infty}(n-2)t^{n-3} a_n(z).
\end{equation}
Comparison of (\ref{eq:dk}) and (\ref{eq:asymptoticda})
demonstrates that the DeWitt-Seeley coefficients other than $a_2$
are expressible in the desired form.

It is convenient to introduce a new set of coefficients by writing
$\lim_{z'\rightarrow z} \cd^2\ex^{t\Box_+}\d_{+}(z,z')$ as an
asymptotic series,
\begin{equation}\label{eq:chiralkernelpowerseries}
\lim_{z'\rightarrow z} \cd^2\ex^{t\Box_+}\d_{+}(z,z') =\frac{1}{16
\p^2 t^2}\sum_{n=0}^{\infty}t^{n} c_n(z),
\end{equation}
where simple computation reveals
\begin{equation}
c_0=-4 \, \mathbb{I}, \qquad c_1=0,
\end{equation}
whilst comparison of (\ref{eq:dk}), (\ref{eq:asymptoticda}) and
(\ref{eq:chiralkernelpowerseries}) yields
\begin{equation}\label{eq:coeffidenitity}
a_n(z)=\frac{1}{16(n-2)}\ba{\cd}^2(c_{n-1}(z)) \qquad n\geq3.
\end{equation}
The effective action can then be written as
\begin{multline}
\qquad\qquad \G^{(1)}[V] =  \frac{1}{16 \p^2}
\ln{\left(\frac{\m}{m}\right)}\inte^{6}z \,
\tr_{\mathcal{R}}(W^2)\\
    - \frac{1}{128 \p^2} \sum_{n=3}^\infty \frac{(n-3)!}{(n-2)
m^{2n-4}}\inte^{8}z \, \tr_{\mathcal{R}}(c_{n-1}),
\end{multline}
the second term now being expressed in full superspace.

Consequently, determining the effective action reduces to
computing the new coefficients $c_n$,  which can of course be
obtained by the same techniques used for computing
DeWitt-Seeley coefficients. If desired, the DeWitt-Seeley
coefficients themselves can be recovered through identity
(\ref{eq:coeffidenitity}), which is nothing more than a projection
onto the chiral subspace.

\setcounter{equation}{0}
\section{The method of computation \label{sec:methodofcomputation}}
We wish to compute the effective action to order $F^5$, which
corresponds to evaluating $c_2$, $c_3$ and $c_4$.  The traditional
method for computing heat kernel coefficients is the iterative
DeWitt method (for example see \cite{kuzenko}).   We however
prefer to use the method developed in \cite{mcarthur,gargett} and
modified for non-Abelian backgrounds in \cite{grasso} since this
calculation parallels that of \cite{grasso}, with only minor
modifications for application to chiral subspace.

We begin by introducing a plane wave basis for the chiral delta
function\footnote{From here onward we work in the chiral
representation.},
\begin{gather}
\d_{+}(z,z')=4 \mathbb{I} \int\!\!\frac{\dd^{4}k}{(2 \p)^4}\;
\ex^{\ii k^a \w_a} \int\!\!\dd^{2} \e \;e^{\ii \e^\a(\q-\q')_\a}
\end{gather}
where\footnote{Note that although $\w_a$ is not itself chiral,
$\ba{\cd}_{\.{\a}}(\w_a)=-\ii(\s_a)_{\a\.{\a}}(\q-\q')^\a$, the
entire delta function is annihilated by $\ba{\cd}_{\.\a}$ since
$(\q-\q')^3=0$.}
\begin{equation} \w_a=x_a-x'_{a}-i \q \s_a
\ba{\q}'+i \q' \s_a \ba{\q}.
\end{equation}
In the coincidence limit, $\cd^2\ex^{t\Box_+}\d_{+}(z,z')$ becomes
\begin{equation}
\lim_{z'\rightarrow z} \cd^2\ex^{t\Box_+}\d_{+}(z,z')=
K^{\a}_{\phantom{\a} \a}(z,t)=\inte \h_+ \;X^{\a}X_{\a}\ex^{t \D},
\end{equation}
the $X$'s being defined by
\begin{align}
X_a&=\cd_a+\ii k_a \\
X_{\a}&=\cd_{\a}+\ii\e_\a,
\end{align}
and where the notation
\begin{gather}\label{eq:moments}
K_{A_1 A_2 \ldots A_n}(z,t)=\inte \h_+ \; X_{A_1}X_{A_2}\ldots
X_{A_n} e^{t\D} \\ \label{eq:measure} \inte \h_+ = 4
\int\!\!\frac{\dd^{4}k}{(2 \p)^4}\int\!\!\dd^{2}\e \\
\D=X^a X_a-W^{\a}X_{\a}
\end{gather}
has been introduced. Note that there is also a shift
$-k_{\a\.{\a}}(\ba{\q}-\ba{\q}')^{\.{\a}}$ in $\cd_{\a}$ which
always vanishes in the coincidence limit since there are no
$\ba{\cd}_{\.{\a}}$ operators present. The X's satisfy the algebra
\begin{equation}\label{eq:Xalgebra}
\{X_\a,X_\b\}=0,\qquad [X_a,X_b]=G_{ab}, \qquad [X_\a,X_{a}]=\ii
(\s_a)_{\a \.{\a}}\ba{W}^{\.{\a}}.
\end{equation}
In this notation the power series
(\ref{eq:chiralkernelpowerseries}) is:
\begin{equation}\label{eq:asymptoticb}
K^{\a}_{\phantom{\a}\a}(z,t) =\frac{1}{16 \p^2
t^2}\sum_{n=0}^{\infty}t^{n} c_n (z).
\end{equation}

Differentiating $K^{\a}_{\phantom{\a}\a}(z,t)$ with respect to $t$
yields the differential equation
\begin{equation}\label{eq:de1}
\frac{d K^{\a}_{\phantom{\a}\a}(z,t)}{d t}= K^{\a
\phantom{\a}a}_{\phantom{\a} \a \phantom{\a}a}(z,t).
\end{equation}
Using the identities
\begin{equation}\label{eq:id1}
0=\inte\h_+ \; \frac{\pd \phantom{k_b}}{\pd k_b}\left(X_\a X_\b
X_a \ex^{t\D}\right)
\end{equation}
and
\begin{equation}\label{eq:id2}
[A, \ex^B]=\int_{0}^{1}\!\!\dd s\;\ex^{s B}[A,B]\ex^{(1-s)B},
\end{equation}
it follows that
\begin{equation}
0=\ii \d_a^b K_{\a\b}(z,t)+2 \ii
t\inte\h_+\;X_{\a}X_{\b}X_a\sum_{n=0}^{\infty}\frac{t^n}{(n+1)!}
ad_{\D}^{\, n}(X^b) \;\ex^{t\D}
\end{equation}
where $ad^{n}$ denotes $n$ nested commutators:
\begin{equation}
ad_{A}^{\, 0}(B)=B, \qquad
ad_{A}^{\,n}(B)=[A,\,ad_{A}^{\,n-1}(B)].
\end{equation}
After contraction of indices, this becomes
\begin{equation}
K^{\a \phantom{\a}a}_{\phantom{\a} \a \phantom{\a}a}(z,t)
=-\frac{2}{t}K^{\a}_{\phantom{\a}\a}(z,t)
-\inte\h_+\;X^{\a}X_{\a}X^a\sum_{n=1}^{\infty}\frac{t^n}{(n+1)!}
ad_{\D}^{\, n}(X_a) \;\ex^{t\D}.
\end{equation}

Inserting this into the differential equation (\ref{eq:de1}), one
obtains:
\begin{equation}\label{eq:de2}
\frac{d K^{\a}_{\phantom{\a}\a}(z,t)}{d
t}+\frac{2}{t}K^{\a}_{\phantom{\a}\a}(z,t)=
-\inte\h_+\;X^{\a}X_{\a}X^a\sum_{n=1}^{\infty}\frac{t^n}{(n+1)!}
ad_{\D}^{\, n}(X_a) \;\ex^{t\D},
\end{equation}
the significance of which is seen in terms of the expansion
(\ref{eq:asymptoticb}), where the left hand side is
\begin{equation}\label{eq:powerseries}
\frac{d K^{\a}_{\phantom{\a}\a}(z,t)}{d
t}+\frac{2}{t}K^{\a}_{\phantom{\a}\a}(z,t)=\frac{1}{16
\p^2}\sum_{n=0}^{\infty}n t^{n-3} c_n (z) = \frac{c_1 (z)}{16 \p^2
t^2}+\frac{2 c_2 (z)}{16 \p^2 t}+\frac{3 c_3 (z)}{16 \p^2}+\ldots
\end{equation}
As is usually the case with this approach, the differential
equation yields an expansion where the first non-trivial
coefficient $c_0$ is absent.  The objective now becomes to
determine the coefficients $c_n (z)$ by expanding the right hand
side of (\ref{eq:de2}) in a power series in $t$, and identifying
it with the right hand side of (\ref{eq:powerseries}).

Since the summation in (\ref{eq:de2}) involves the repetitive
calculation of commutators, it is first useful to establish the
following relations:
\begin{align}
\lbrack\D,X_a\rbrack &= 2 G^b_{\phantom{b}a}X_b + (\cd_a
W^\a)X_\a+\ii (\s_a)_{\a\.{\a}}\ba{W}^{\.{\a}}W^{\a}\no \\
\lbrack \D,X_\a \rbrack &=(\cd_\a W^\b)X_\b
-2\ii(\s^a)_{\a\.{\a}}\ba{W}^{\.{\a}}X_a \label{eq:commutation2}
\\\lbrack \D,A\rbrack &=(\cd^a\cd_a A)+2(\cd^a A)X_a-W^\a(\cd_\a
A)-(-1)^{\ce(A)}\gc{W^\a}{A}X_\a.\no
\end{align}

From these it is clear that summation will generate a series of
moments of the form $K_{A_{1} \ldots A_{i}}(z,t)$ as defined in
(\ref{eq:moments}). Furthermore, it is not difficult to show that
to order $n$ in this summation, the moments generated have at most
$(n+3)$ indices. It is convenient to always place these indices in
a specific order: first spinor, then vector. This can be achieved
through the commutation relations (\ref{eq:commutation}). With
such an ordering, the leading term in a moment's asymptotic power
series has the following behaviour\footnote{By `leading term' we
mean the first (expected) non-trivial term, ie $K_+(z,t)$ has a
leading term of order $t^0$.}:
\begin{equation}\label{eq:tbehaviour}
K_{A_1 \ldots A_{q+p}}(z,t)\sim \frac{1}{t^2}
\left(\frac{1}{t}\right)^{\left[\frac{p}{2}\right]}t^{2-q} =
t^{-q-\left[\frac{p}{2}\right]}\qquad \qquad q\leq2
\end{equation}
where $K_{A_1 \ldots A_{q+p}}(z,t)$ has $q$ undotted spinor
indices, $p$ vector indices and $[\frac{p}{2}]$ denotes the
largest integer part of $\frac{p}{2}$. Moments with greater than
two undotted spinor indices vanish since $X_\a X_\b X_\g=0$.

From these considerations, and by comparison with
(\ref{eq:powerseries}), the summation in equation (\ref{eq:de2})
truncates at $n=2k-1$ when evaluating $c_k (z)$ for $k\geq2$.
Moreover, it turns out the last term in this truncated summation
always vanishes due to the fact that it takes the form
\begin{equation}
-\frac{(2t)^{2k-1}}{(2k)!}(\cd^{a_1}\cd^{a_2}\ldots
\cd^{a_{2k-2}}G^{a_{2k-1}a_{2k}}) K^{\a}_{\phantom{\a}\a a_1 a_2
\ldots a_{2k}}(z,t)\qquad \qquad k\geq2
\end{equation}
and the moment is only ever required to leading order in its power
series in $t$.  To this order the moment is always totally
symmetric in its spacetime indices, whereas $G$ is antisymmetric.
Consequently all such terms vanish\footnote{Alternatively, to this
order the moment is proportional to the identity matrix in its
group indices, and the coefficient therefore vanishes under
integration by parts.}, and when evaluating $c_k(z)$ the summation
truncates at $n=2k-2$.

To compute $c_k(z)$ one must expand the set of moments which
result from the summation in (\ref{eq:de2}) to appropriate order
in $t$. This is achieved through either direct expansion of the
moment's exponential, or iteratively through the use of the
identities
\begin{align}\label{eq:idzerok}
0=&\inte\h_+ \; \frac{\pd \phantom{k_b}}{\pd k_b}\left(X_{A_1}
\ldots X_{A_n} \ex^{t\D}\right) , \\
\label{eq:idzeroe} 0=&\inte\h_+ \; \frac{\pd \phantom{\e_\a}}{\pd
\e_\a}\left(X_{A_1} \ldots X_{A_n} \ex^{t\D}\right)
\end{align}
and
\begin{equation}\label{eq:idzerode}
\frac{d^m K_{A_1 \ldots A_n}(z,t)}{d t^m}=\inte \h_+ \;
X_{A_1}\ldots X_{A_n} \D^m e^{t\D} .
\end{equation}
These can be used to express the desired moment in terms of
moments which are easier to compute. For greater detail we direct
the reader Appendix A which illustrates this procedure, or to the
earlier work \cite{grasso}.

\setcounter{equation}{0}
\section{Results} \label{sec:results}

\subsection{Superfield form}

The results produced directly by the calculational procedure
employed here can always be vastly simplified since almost all
possible combinations of covariant derivatives on field strengths
are produced. Surprisingly, there are so many terms generated in
$c_4$ that it is no longer practical to compute by hand. All
results can be brought into their most compact form through
integration by parts, the cyclic property of the trace, the
equations of motion, the repetitive use of the commutation
relations (\ref{eq:commutation}), and application of on-shell
identities such as
\begin{gather}
\cd_\a \cd_\b \ba{\cd}_{\.{\a}}\ba{W}_{\.{\b}}=4\ce_{\a\b}\{
\ba{W}_{\.{\a}},\ba{W}_{\.{\b}} \},  \qquad \qquad \cd_\a
\ba{\cd}_{\.{\a}}\cd_\b W^\a=- 4\{ \ba{W}_{\.{\a}},W_\b \} ,\\
(\cd^a \cd_a W^\a)=[W^\b,\cd_\b W^\a],\label{eq:ddw}
\end{gather}
the latter being established through equation (\ref{eq:iddd}).

Despite the eventual simplicity of the result, the computation of
$c_4$ is very involved and requires a great deal of work. Unlike
$c_2$ and $c_3$, $c_4$ is not manifestly real after
simplification, but can be brought into such a form by using the
identities (modulo integration by parts):
\begin{gather}
\tr_{\mathcal{R}}\Big((\cd^a \1)(\cd_a \3)\4\2\Big) =
\tr_{\mathcal{R}}\Big((\cd^a \3)(\cd_a \1)\2\4\Big), \\
\begin{split}
\tr_{\mathcal{R}}\Big((\cd^a \1)(\cd_a \2)\6+2& (\cd_\a
\1[\b])\1\2[\b]\6\Big) \\ &=\tr_{\mathcal{R}}\Big((\cd^a \3)(\cd_a
\4)\5+2 (\ba{\cd}_{\.{\a}}\4[\b])\4\3[\b]\5\Big).
\end{split}
\end{gather}

Finally, to this order the one-loop effective action is computed
to be:
\begin{multline}\label{eq:result}
\G^{(1)}_{\c,\mathcal{R}}=  \frac{1}{16 \p^2}
\ln{\left(\frac{\m}{m}\right)}\inte^{6}z \, \tr_{\mathcal{R}}(W^2)
- \frac{1}{128 \p^2 m^{2}}\inte^{8}z \, \tr_{\mathcal{R}}(c_{2})
\\- \frac{1}{256 \p^2 m^{4}}\inte^{8}z \, \tr_{\mathcal{R}}(c_{3})
- \frac{1}{192 \p^2 m^{6}}\inte^{8}z \, \tr_{\mathcal{R}}(c_{4})
\end{multline}
where
\begin{align}
\label{eq:c2}
\tr_{\mathcal{R}}(c_2)=&\frac{1}{3}\tr_{\mathcal{R}}(G^{ab}G_{ba})=-\frac{1}{6}\tr_{\mathcal{R}}\left(
(\ba{\cd}_{\.{\a}}\ba{W}_{\.{\b}})(\ba{\cd}^{\.{\b}}\ba{W}^{\.{\a}})+
(\cd^{\a}W^{\b})(\cd_{\b}W_{\a})\right)
\\ \no \\
\label{eq:c3}
\tr_{\mathcal{R}}(c_3)=&\frac{2}{15}\tr_{\mathcal{R}}(\1\3\2\4-4
\5\6)
\\ \no \\
\begin{split} \label{eq:c4}
\tr_{\mathcal{R}}(c_4)=&\frac{1}{105}\tr_{\mathcal{R}}\Big(2(\cd^a
\1)(\cd_a \3)\2\4 -6(\cd^a \1)(\cd_a \2)\6 \\& \qquad\qquad
-3(\cd^a \1)(\cd_a \3)\4\2 + 18 (\cd_\a \1[\b])\1\2[\b]\6
\\&\qquad\qquad\qquad\qquad\qquad\qquad\qquad\qquad
+\frac{5}{2}(\cd_\a \1[\b])\1\3\2[\b]\4 \Big)+c.c.
\end{split}
\end{align}
and c.c. denotes the complex conjugate.

Note that $c_2$ actually vanishes under integration by parts since
on-shell $(\cd_\a \cd_\b W_\g) =0$, and therefore provides no
contribution to the effective action.

\subsection{Component form}

The component form of the above expressions can be extracted
through the usual techniques (for example see \cite{kuzenko}). The
the bosonic component of $c_3(z)$ is computed to be:
\begin{equation}
\frac{1}{30}\tr_{\mathcal{R}}(2\FAA +3\FBB-4\FCC-16\FEE),
\end{equation}
where
\begin{equation}
\begin{split}
\nab_a=\pd_a-\ii A_a, \qquad \qquad \qquad  [\nab_a,\nab_b]=-\ii
F_{a b} \qquad \qquad \\
F_{ab}=\pd_a A_b-\pd_b A_a-\ii [A_a,A_b], \qquad
\nab_c(F_{ab})=\pd_{c}F_{ab}-\ii [A_c,F_{ab}].
\end{split}
\end{equation}
As will be shown, this agrees exactly with the component results
given in \cite{vdv} and \cite{mt}.

Extraction and comparison of the bosonic component of $c_4(z)$ is
complicated by the fact that in the non-Abelian case there are
many possible tensor structures which are not all independent.
More generally, in addition to some identities which are dependent
on the spacetime dimension, structures with a given number of
contracted $F$'s and $\nab F$'s may be related through: the
Bianchi identity, the equations of motion, integration by parts,
the cyclic property of the trace over the gauge indices, and the
non-Abelian identity
\begin{equation}
[F_{ab},F_{cd}]=2 \ii \nab_{[a}\nab_{b]}F_{cd}.
\end{equation}
An independent set of such tensor structures forms a basis, and
different bases are used throughout the literature, since
different calculational procedures naturally select different
bases. In the case presented here the basis is almost completely
determined by the use of superspace.  In particular, in superspace
and hence at the component level, both covariant derivatives act
on adjacent field strengths and are always contracted with one
another.  Complete details regarding the basis used here and the
transformation from any basis into it are given in Appendix B.

The bosonic component of $c_4(z)$ in this basis is:
\begin{multline}\label{eq:c4component}
-\frac{1}{210}\tr_{\mathcal{R}}\Big(19\Fa+11\Fb+13\Fc
\\+32\Fd-60\Fe+\frac{261\ii}{5}\FA\\
-89\ii \FB -41\ii\FC-\frac{7\ii}{5}\FD\Big).
\end{multline}

\setcounter{equation}{0}
\section{Comparison with literature} \label{sec:comparison}
To date the bosonic DeWitt-Seeley coefficients associated with
scalars, vectors or spinors in the presence of non-Abelian
background Yang-Mills fields in arbitrary spacetime dimension have
been separately computed to low order \cite{vdv,mt,ft}. Since at
the component level the action (\ref{eq:action}) corresponds to
supersymmetric matter (a set of massive scalars and their
fermionic superpartners) coupled to a background non-Abelian
supersymmetric Yang-Mills field, a non-trivial check of the
results derived here is available.

From the tables in \cite{mt} and \cite{ft}, one can assemble the
total bosonic component of the DeWitt-Seeley coefficients
associated with a theory possessing $N_1$ vectors, $N_0$ scalars
and $N_{1/2}$ spinors all in the adjoint representation, coupled
to a Yang-Mills background by using\footnote{In the notation used
in \cite{mt} and \cite{ft}, $a_n=b_{2n}$}
\begin{equation}\label{eq:sdw}
a^{\mathrm{tot}}_{n}=N_1 a_n(\D_1)+(N_0-2
N_1)a_n(\D_0)-\frac{N_{1/2}}{\g}a_n(\D_{1/2})
\end{equation}
where $\g=1,2,4$ for Dirac, Majorana and Majorana-Weyl spinors
respectively, and $a_n(\D_s)$ ($s=0,1,\frac{1}{2}$) denotes the
contribution generated by the presence of second order (scalar,
vector, spinor) operators in the original action.

At the component level the starting action (\ref{eq:action})
contained two scalars and two Majorana-Weyl spinors in $D=4$, so
from \cite{mt,ft} and equation (\ref{eq:sdw}), one generates the
following on-shell bosonic components of the DeWitt-Seeley
coefficients:
\begin{align}\tr_{\Ad}(a_3) & = 0 \\
\begin{split}\tr_{\Ad}(a_4) & =
-\frac{1}{240}\tr_{\Ad}\Big(2\FAA +3\FBB
\\&\qquad\qquad\qquad\qquad\qquad \qquad\qquad-4\FCC-16\FEE \Big)
\end{split} \\
\begin{split}\label{eq:f5frommt}
\tr_{\Ad}(a_{5})&=\frac{1}{21}\frac{1}{5!}\tr_{\Ad}\Big(-10 \Fd
-32\Fe
\\&\qquad\qquad\qquad+8\fd +\frac{1}{2}\Fj\\&\qquad\qquad\qquad-42\Fg+6\fg
\\&\qquad\qquad\qquad +6\Fa-\frac{19}{2}\Fb\\&\qquad\qquad\qquad\qquad\qquad
-28\fk\Big) +(F^5 \; \mathrm{terms})
\end{split}
\end{align}
The vanishing of $a_3$ is non-trivial, and as indicated only the
derivative terms of $a_5$ have so far been computed at the
component level \cite{vdv,mt,ft}. Inspection reveals immediate
agreement, up to an overall numerical multiplicative constant,
between $a_3$, $a_4$ and the bosonic components of $c_2$ and $c_3$
respectively.  Taking into account the relationship between the
coefficients $c_n$ and $a_n$ in superspace, equation
(\ref{eq:coeffidenitity}), and restricting $\mathcal{R}$ to be the
adjoint representation, exact agreement is found.

Comparison between $a_{5}$ and $c_4$ is far less trivial since the
results for $a_5$ in the literature are expressed in a different
basis, which is non-minimal in $D=4$. After expressing $a_5$ in
our basis (see Appendix B), we find:
\begin{multline}\label{eq:f5frommmtbasis}
\tr_{\Ad}(a_{5})=
\frac{1}{21}\frac{1}{5!}\tr_{\Ad}\Big(19\Fa+11\Fb
\\+13\Fc+32\Fd\\-60\Fe\Big)+(F^5 \; \mathrm{terms}).
\end{multline}
Comparing with the bosonic component of $c_4$ in equation
(\ref{eq:c4component}), and again taking into account equation
(\ref{eq:coeffidenitity}), exact agreement of the derivative terms
is found.

\section{Discussion}\label{sec:discussion}
We can now give the ingredients for the calculation of the
one-loop effective action for an arbitrary $\mathcal{N}=2$ SYM
theory to order $F^5$. As previously noted, the effective action
will be given by an appropriate linear combination of: (i)
$\G^{(1)}_{\c,R \oplus R_c}$ and $\G^{(1)}_{\c,\Ad}$ given by
(\ref{eq:result}; (ii) the logarithm of the superdeterminant of
the background covariant operator in the quadratic part of the
action for the vector superfield in the adjoint representation.
The latter,
\begin{equation}
\frac{\ii}{2}\ln\mathrm{sDet(\Box_v-M^2)},
\end{equation}
where $M$ is an infrared regulator and the background vector
d'Alembertian, $\Box_\mathrm{v}$, is
\begin{equation}
\Box_\mathrm{v}=\cd^a\cd_a-\1\cd_\a+\3\ba{\cd}^{\.{\a}},
\end{equation}
was derived in \cite{grasso}, and to order $F^5$ is:
\begin{equation}\label{eq:vea}
\frac{\ii}{2}\ln\mathrm{sDet}(\Box_\mathrm{v}-M^2)= \frac{1}{32
\p^2 M^{4}}\inte^{8}z \, \tr_{\Ad}(a^{\mathrm{(v)}}_4) + \frac{1}{16
\p^2 M^{6}}\inte^{8}z \, \tr_{\Ad}(a^{\mathrm{(v)}}_5)
\end{equation}
where\footnote{Here the expression for
$\tr_{\Ad}(a^{\mathrm{(v)}}_5)$ from \cite{grasso} has been
simplified through use of equation (\ref{eq:ddw}).}
\begin{align}
\tr_{\Ad}(a^{\mathrm{(v)}}_4)=& \frac{1}{3}\, \tr_{\Ad}(2\, \5 \6-
\1
\3 \2 \4) \\
\tr_{\Ad}(a^{\mathrm{(v)}}_5)=& \frac{1}{30}\tr_{\Ad} \big((\cd^a
\1)(\cd_a \2) \6+ (\cd^a \1)(\cd_a \3) \4 \2 \no \\&\qquad\qquad-
(\cd^a \1)(\cd_a \3) \2 \4 -3(\cd_\a W^\b) \1 W_\b \6 \no
\\&\qquad\qquad\qquad -(\cd_\a W^\b) \1 \3 W_\b \4\big)+ c.c.
\end{align}

Finally, it is worth pointing out that in \cite{grasso}, when
computing the one-loop effective action for $\mathcal{N}=4$ SYM
(which is just (\ref{eq:vea})), we where able to push this method
for calculating heat kernel coefficients to order $F^6$, whereas
in the chiral case presented here, only to order $F^5$ with
comparable difficulty. This is primarily due to the fact that in
the $\mathcal{N}=4$ case the associated heat kernel has the power
series behaviour,
\begin{equation}
K_{\mathrm{v}}(z,t)=\frac{t^2 a_4^{\mathrm{(v)}}}{16 \p^2} +
\frac{t^3 a_5^{\mathrm{(v)}}}{16 \p^2}+ \ldots,
\end{equation}
whereas the power series of the chiral kernel was given by
equation (\ref{eq:asymptotica}).  Consequently computing $F^5$
terms in $\mathcal{N}=4$ SYM (ie $a_5^{\mathrm{(v)}}$), merely
involved the computation of the second non-trivial coefficient,
whereas one must compute the forth non-trivial coefficient to
acquire the $F^5$ contributions in the chiral case.

\section*{Acknowledgements}
I wish to thank S.M. Kuzenko for the suggestion of this project,
and both S.M. Kuzenko and I.N. McArthur for suggestions,
discussions and references.

\setcounter{equation}{0}
\appendix
\section{The method of computation}
Here we illustrate through example the method of computation of
coincidence limits of moments of heat kernels employed in this
paper.

The power series arguments presented in Section
\ref{sec:methodofcomputation} showed that in computing $c_2(z)$
the summation in equation (\ref{eq:de2}) truncates at $n=2$.
Consider the $n=1$ contribution:
\begin{equation}
-\frac{t}{2!}\inte\h_+\;X^{\a}X_{\a}X^a [\D,X_a] \;\ex^{t\D}.
\end{equation}
Using the commutation relations (\ref{eq:Xalgebra}) and
(\ref{eq:commutation2}) this reduces to
\begin{equation}\label{eq:kernelexpansion}
-\frac{t}{2!}\Big(Q^{\a\b a}K_{\a \b a}(z,t)+R^{\a a b}K_{\a a
b}(z,t)+S^{\a \b }K_{\a \b}(z,t)\Big),
\end{equation}
where only the moments which contribute to $c_2(z)$ have been
retained\footnote{For example from (\ref{eq:powerseries}) and
(\ref{eq:tbehaviour}), one observes that terms like $t K(z,t)$, $t
K_\a(z,t)$ and $t K_{a b}(z,t)$ etc will provide no contribution
to $c_2(z)$.} and on-shell
\begin{align}
Q^{\a\b a}&=\ii \ce^{\b\a} (\s^a)_{\g\.{\g}} \4[\g] \1[\g] -2
(\cd^\a\cd^a\1[\b])+ 2 \ce^{\b\a} (\cd^b G^a_{\phantom{a}b})\\
R^{\a a b}&=4 (\cd^\a G^{b a})\\ S^{\a \b}&=-2(\cd^\a \cd^a \cd_a
\1[\b])+\ce^{\b\a}G^{ba}G_{ab}\\ G_{ab}&=[X_a,X_b]=[\cd_a,\cd_b].
\end{align}
The remaining moments in (\ref{eq:kernelexpansion}) must now be
expanded in a power series to the required order in $t$. In the
current example this does not pose any additional difficulties
since all moments are required only to leading order (as in
equation (\ref{eq:tbehaviour})), which can easily be obtained
simply by expanding the exponential in the moment into
commutators.  For example:
\begin{equation}
K_{\a a b}(z,t)=\inte \h_+ \; X_\a X_a X_b e^{t\D}=\frac{1}{16
\p^2}\frac{1}{t^2}\h_{ab}W_\a+\mathcal{O}(t^{-1}),
\end{equation}
where the $k$ integral has been performed (after Wick rotation to
a Euclidean metric).

However, if we wish to compute $c_3(z)$, such a moment would have
been required to subleading order.  In such a case one does not
simply expand the exponential into commutators, since this is a
very cumbersome method to evaluate most moments to other than
leading order. Rather, it can be expressed in terms of moments
which are easier to compute, or need only be computed to first
order. Here this can be achieved through the use of the identity
(\ref{eq:idzeroe}), in the form:
\begin{equation}
0=\inte\h_+ \; \frac{\pd \phantom{\e_\a}}{\pd \e_\g}\left(X_{\a}
X_{\b}X_{a}X_{b} \ex^{t\D}\right).
\end{equation}
After contraction of $\g$ and $\b$, and using equation
(\ref{eq:id2}), this leads to an expression for $K_{\a a b}(z,t)$,
\begin{equation}
K_{\a a b}(z,t)= t\inte\h_+\;X_{\a}X_{\b}X_a X_b
\sum_{n=0}^{\infty}\frac{t^n}{(n+1)!} ad_{\D}^{\, n}(W^\b)
\;\ex^{t\D}.
\end{equation}
Examining the power series behaviour it can be seen that to
compute $K_{\a a b}(z,t)$ to subleading order, this summation will
truncate at $n=2$, but in doing so generates another series of
moments which also need to be computed, ie
\begin{equation}\label{eq:kaab}
K_{\a a b}(z,t)= t W^\b K_{\a\b a b}(z,t) + \ldots
\end{equation}
In turn these moments may be expressed in terms of others using
identifies like (\ref{eq:idzerok}), (\ref{eq:idzeroe}) or
(\ref{eq:idzerode}).

Proceeding iteratively in this fashion, when computing $c_k(z)$
one eventually establishes a small group of moments which need to
be computed to particular order.  This group of moments naturally
arrange themselves into a hierarchy, where computing any moment in
the hierarchy depends on the computation of other moments higher
up in the hierarchy (for example $K_{\a a b}(z,t)$ depends on
$K_{\a\b a b}(z,t)$ in (\ref{eq:kaab})).  One can then proceed to
compute, in systematic fashion, all moments within the hierarchy
by starting at the top.

\setcounter{equation}{0}
\section{Change of basis}
In this appendix we briefly describe the basis used in this paper,
and outline the general transformation from any other basis into
it. In particular the expression for $a_{5}$ in equation
(\ref{eq:f5frommt}), is transformed into
(\ref{eq:f5frommmtbasis}).

For simplicity the following notation is introduced (where the
trace is over the gauge indices in some representation
$\mathcal{R}$ of the gauge group):
\begin{align*}
s_{0,0}&=\tr_{\mathcal{R}}(\FA) & s_{0,1}&=\tr_{\mathcal{R}}(\FB) \\
s_{0,2}&=\tr_{\mathcal{R}}(\FC) & s_{0,3}&=\tr_{\mathcal{R}}(\FD)\\
s_{0,4}&=\tr_{\mathcal{R}}(\FE) & s_{0,5}&=\tr_{\mathcal{R}}(\FF) \\
s_{1,0}&=\tr_{\mathcal{R}}(\Fa) & s_{1,1}&=\tr_{\mathcal{R}}(\Fb) \\
s_{1,2}&=\tr_{\mathcal{R}}(\Fc) & s_{1,3}&=\tr_{\mathcal{R}}(\Fd) \\
s_{1,4}&=\tr_{\mathcal{R}}(\Fe) & s_{1,5}&=\tr_{\mathcal{R}}(\Ff) \\
s_{1,6}&=\tr_{\mathcal{R}}(\Fg) & s_{1,7}&=\tr_{\mathcal{R}}(\Fh) \\
s_{1,8}&=\tr_{\mathcal{R}}(\Fi) & s_{1,9}&=\tr_{\mathcal{R}}(\Fj) \\
s_{1,10}&=\tr_{\mathcal{R}}(\Fk) & s_{1,11}&=\tr_{\mathcal{R}}(\Fl) \\
s_{2,1}&=\tr_{\mathcal{R}}(\fb) &
s_{2,3}&=\tr_{\mathcal{R}}(\fd) \\
s_{2,4}&=\tr_{\mathcal{R}}(\fe) & s_{2,5}&=\tr_{\mathcal{R}}(\ff) \\
s_{2,6}&=\tr_{\mathcal{R}}(\fg) & s_{2,7}&=\tr_{\mathcal{R}}(\fh) \\
s_{2,9}&=\tr_{\mathcal{R}}(\fj) & s_{2,10}&=\tr_{\mathcal{R}}(\fk) \\
s_{2,11}&=\tr_{\mathcal{R}}(\fl) & s_{2,12}&=\tr_{\mathcal{R}}(\fm) \\
s_{2,13}&=\tr_{\mathcal{R}}(\fn) & s_{2,14}&=\tr_{\mathcal{R}}(\fo) \\
s_{2,15}&=\tr_{\mathcal{R}}(\fp) & s_{2,16}&=\tr_{\mathcal{R}}(\fq) \\
s_{2,17}&=\tr_{\mathcal{R}}(\fr) &
s_{2,18}&=\tr_{\mathcal{R}}(\fs).
\end{align*}
This is not a complete list, and excludes terms which can
obviously be reduced to a linear combination of the above (modulo
integration by parts and the equations of motion). In $D=4$ the
following set provides a basis for all such possible tensor
structures:
\begin{equation}\label{eq:basis}
\{s_{0,0},s_{0,1},s_{0,2},s_{0,3},s_{1,0},s_{1,1},
s_{1,2},s_{1,3},s_{1,4},s_{2,3}\},
\end{equation}
which consists of four $F^5$ structures, five structures with
contracted covariant derivatives acting on adjacent field
strengths, and a single structure with two covariant derivatives
which are not contracted.  We demonstrate this below.

\subsection{$F^5$ terms}
One can readily establish that the six $F^5$ terms: $s_{0,0}$,
$s_{0,1}$, $s_{0,2}$, $s_{0,3}$, $s_{0,4}$ and $s_{0,5}$, are not
linearly independent in four dimensions by using the following
$\s$ matrix identity:
\begin{equation}\label{eq:sigmaidentity}
\mathrm{tr}(\s^a \~\s^b \s^c \~\s^d \s^e \~\s^f) =\mathrm{tr}(\s^b
\~\s^a \s^f \~\s^e \s^d \~\s^c).
\end{equation}
This implies
\begin{equation}
\Big(\mathrm{tr}(\s^a \~\s^b \s^c \~\s^d \s^e
\~\s^f)\mathrm{tr}(\~\s^g \s^h \~\s^i
\s^j)+c.c\Big)\tr_{\mathcal{R}}(F_{ab}F_{cd}F_{ef}F_{gh}F_{ij}
+F_{ab}F_{ef}F_{cd}F_{gh}F_{ij})=0,
\end{equation}
which reduces to
\begin{equation}\label{eq:f51}
s_{0,1}-2 s_{0,2} + s_{0,3}+\frac{1}{2} s_{0,4}+\frac{3}{2}
s_{0,5}=0.
\end{equation}
Similarly one can use (\ref{eq:sigmaidentity}) to establish
\begin{equation}
\Big(\mathrm{tr}(\s^a \~\s^b \s^c \~\s^d \s^e
\~\s^f)\mathrm{tr}(\~\s^g \s^h \~\s^i
\s^j)+c.c\Big)\tr_{\mathcal{R}}(F_{ab}F_{cd}F_{gh}F_{ef}F_{ij}
+F_{ab}F_{ef}F_{gh}F_{cd}F_{ij})=0,
\end{equation}
which becomes
\begin{equation}
\label{eq:f52}  s_{0,0}-2 s_{0,1}- s_{0,2}+ \frac{3}{2}
s_{0,4}-\frac{1}{2} s_{0,5}=0.
\end{equation}
Equations (\ref{eq:f51}) and (\ref{eq:f52}) then allow two of the
six $F^5$ structures to be expressed in terms of the other four.
For example we choose to treat $s_{0,4}$ and $s_{0,5}$ as
dependent:
\begin{align}
s_{0,4}=&-\frac{3}{5}s_{0,0}+s_{0,1}+s_{0,2}+\frac{1}{5}s_{0,3}\\
s_{0,5}=& \frac{1}{5}s_{0,0}-s_{0,1}+s_{0,2}+\frac{3}{5}s_{0,3}.
\end{align}

\subsection{$D^2 F^4$ terms}
Using the equations of motion, the Bianchi identity, integration
by parts, the cyclic property of the trace and the identity
\begin{equation}\label{eq:nonab}
[F_{ab},F_{cd}]=2 \ii \nab_{[a}\nab_{b]}F_{cd},
\end{equation}
each of the terms $s_{1,i}$ with $6\leq i \leq 11$ can be
expressed in the proposed basis:
\begin{align}
s_{1,6}=&-2 \ii s_{0,1}+2 \ii s_{0,2}-s_{1,4}-s_{1,5}\\
s_{1,7}=& 4\ii s_{0,5}-2 s_{1,1}=  \frac{4 \ii}{5}s_{0,0}-4 \ii
s_{0,1}+4\ii s_{0,2}+\frac{12\ii}{5}s_{0,3} -2 s_{1,1}\\
s_{1,8}=&-2 \ii s_{0,0}+2 \ii s_{0,1}-2s_{1,3}\\
s_{1,9}=&4\ii s_{0,4}- s_{1,0}- s_{1,2}=-\frac{12\ii}{5}s_{0,0}
+4\ii s_{0,1}+4 \ii s_{0,2}+\frac{4\ii}{5}s_{0,3}- s_{1,0}- s_{1,2} \\
s_{1,10}=&4\ii s_{0,4}- s_{1,0}- s_{1,2}=-\frac{12\ii}{5}s_{0,0}
+4\ii s_{0,1}+4 \ii s_{0,2}+\frac{4\ii}{5}s_{0,3}- s_{1,0}- s_{1,2} \\
s_{1,11}=&-2 \ii s_{0,1}+2 \ii s_{0,2}-s_{1,4}-s_{1,5}.
\end{align}

\subsection{$D_a D_b F^4$ terms}
Again using these properties one can generate the following
independent equations:
\begin{align}
s_{2,1}=&\ii s_{0,2}-\ii
s_{0,3}+\frac{1}{4}s_{1,1}-\frac{1}{2}s_{2,5}-s_{2,9}\\
s_{2,9}=&-\ii s_{0,1}+\ii
s_{0,2}-s_{1,4}-\frac{1}{2}s_{2,3}+\frac{1}{2}s_{2,6}\\
s_{2,4}=&-\ii s_{0,4}+\ii
s_{0,5}+\frac{1}{2}s_{1,2}-\frac{1}{2}s_{1,10}+s_{2,3}\\
s_{2,5}=& s_{2,6}\\
s_{2,7}=& s_{2,9}\\
s_{2,13}=&s_{2,18}=-\ii s_{0,0}+\ii s_{0,1}-s_{1,3}\\
s_{2,12}=&- s_{2,15}\\
s_{2,12}=&- s_{1,5}+s_{2,7}\\
s_{2,10}=&s_{2,14}=-s_{2,11}-s_{2,12}\\
s_{2,10}=&-s_{2,1}+s_{2,9}\\
s_{2,16}=& -s_{2,17}=\frac{\ii}{2} s_{0,4}-\frac{\ii}{2}
s_{0,5}-s_{2,13}+s_{2,14}.
\end{align}
Furthermore, two more independent equations can be produced by
using identity (\ref{eq:sigmaidentity}), for example:
\begin{multline}
\Big(\mathrm{tr}(\s^a \~\s^b \s^c \~\s^d \s^e
\~\s^f)\mathrm{tr}(\~\s^g \s^h \~\s^i
\s^j)+c.c\Big)\\\Big(\tr_{\mathcal{R}}((\nab_b F_{cd})\nab_a
(F_{gh})F_{ij}F_{ef}-(\nab_a F_{fe})\nab_b
(F_{gh})F_{ij}F_{dc})\Big)=0,
\end{multline}
and
\begin{multline}
\Big(\mathrm{tr}(\s^a \~\s^b \s^c \~\s^d \s^e
\~\s^f)\mathrm{tr}(\~\s^g \s^h \~\s^i
\s^j)+c.c\Big)\\\Big(\tr_{\mathcal{R}}((\nab_a F_{gj})\nab_f
(F_{hi})F_{bc}F_{ed}-(\nab_b F_{gj})\nab_c
(F_{hi})F_{af}F_{de})\Big)=0,
\end{multline}
which reduce to
\begin{equation}
s_{1,0}+s_{1,1}-4 s_{1,4}+4 s_{2,1}+2 s_{2,4}=0
\end{equation}
and
\begin{equation}
s_{1,1}-s_{1,2}+4 s_{1,3}-4 s_{1,4}+8 s_{1,5}-2 s_{2,3}+2
s_{2,4}-4 s_{2,9}+4 s_{2,12}-8 s_{2,15}=0
\end{equation}
respectively.

\subsection{The basis}
All tensor structures can now be expressed in the basis
(\ref{eq:basis}). Introducing the condensed notation,
\begin{align*}
\{a,b,& c,d,e,f,g,h,i,j\} \\ & \equiv a s_{0,0}+b s_{0,1}+c
s_{0,2}+d s_{0,3}+e s_{1,0}+f s_{1,1}+g s_{1,2}+h s_{1,3}+i
s_{1,4}+j s_{2,3}
\end{align*}
we list below all terms expressed in this basis:
\begin{center}
$\begin{array}{rrrrrrrrrrrrr}
s_{0,4}=&\{&-\frac{3}{5},& 1,& 1,& \frac{1}{5},& 0,& 0,& 0,& 0,& 0,& 0&\}\\
s_{0,5}=&\{&\frac{1}{5},&-1,& 1,& \frac{3}{5},& 0,& 0,& 0,& 0,& 0,& 0&\}\\
s_{1,5}=&\{&2\ii ,&-5\ii ,& 0,&-\ii,& \frac{3}{4},& \frac{3}{4},&
\frac{3}{4},& 1,&-3,& 0&\}\\
s_{1,6}=s_{1,11}=&\{&-2\ii,& 3\ii,& 2\ii,& \ii,& -\frac{3}{4},&
-\frac{3}{4},& -\frac{3}{4},& -1,& 2,& 0&\}\\
s_{1,7}=&\{&\frac{4\ii}{5},&-4\ii,& 4\ii,& \frac{12\ii}{5},& 0,&
-2,&
0,& 0,& 0,& 0&\}\\
s_{1,8}=&\{&-2\ii ,& 2\ii,& 0,& 0,& 0,& 0,& 0,& -2,& 0,& 0&\}\\
s_{1,9}=s_{1,10}=&\{&-\frac{12\ii}{5} ,& 4\ii,& 4\ii,&
\frac{4\ii}{5},&-1,& 0,&-1,& 0,& 0,& 0&\}\\
s_{2,1}=&\{&-\ii,&2\ii,&\ii,&0,&-\frac{1}{2},&-\frac{1}{4},&-\frac{1}{2},&0,&1,&-\frac{1}{2}&\}\\
s_{2,4}=&\{&2\ii,&-4\ii,&-2\ii,&0,&\frac{1}{2},&0,&1,&0,&0,&1&\} \\
s_{2,5}=s_{2,6}=&\{&\ii,&-\ii,&-\ii,&-\ii,&\frac{1}{2},&\frac{1}{2},&\frac{1}{2},&0,&0,&1&\}
\\
s_{2,7}=s_{2,9}=&\{&\frac{\ii}{2},&-\frac{3\ii}{2},&\frac{\ii}{2},&-\frac{\ii}{2},&\frac{1}{4},&\frac{1}{4},&\frac{1}{4},&0,&-1,&0&\}
\\
s_{2,10}=s_{2,14}=&\{&\frac{3\ii}{2},&-\frac{7\ii}{2},&-\frac{\ii}{2},&-\frac{\ii}{2},&\frac{3}{4},&\frac{1}{2},&\frac{3}{4},&0,&-2,&\frac{1}{2}&\}
\\
s_{2,11}=&\{&0,&0,&0,&0,&-\frac{1}{4},&0,&-\frac{1}{4},&1,&0,&-\frac{1}{2}&\}
\\
s_{2,12}=-s_{2,15}=&\{&-\frac{3\ii}{2},&\frac{7\ii}{2},&\frac{\ii}{2},&\frac{\ii}{2},&-\frac{1}{2},&-\frac{1}{2},&-\frac{1}{2},&-1,&2,&0&\}
\\
s_{2,13}=s_{2,18}=&\{&-\ii,&\ii,&0,&0,&0,&0,&0,&-1,&0,&0&\} \\
s_{2,16}=-s_{2,17}=&\{&\frac{21\ii}{10},&-\frac{7\ii}{2},&-\frac{\ii}{2},&-\frac{7\ii}{10},&\frac{3}{4},&\frac{1}{2},&\frac{3}{4},&1,&-2,&\frac{1}{2}&\}
\end{array}$
\end{center}
Finally, in this basis we find that the derivative terms of
$a_{5}$, given in equation (\ref{eq:f5frommt}), become
\begin{equation}
a_{5}=\frac{1}{21}\frac{1}{5!}\{\frac{234\ii}{5}, -32\ii,-74\ii,
-\frac{168\ii}{5},19,11,13,32,-60,0\},
\end{equation}
which, ignoring $F^5$ terms, is equation
(\ref{eq:f5frommmtbasis}).

\end{document}